# Presenting the Cyclotactor Project

Staas de Jong
LIACS, Leiden University
staas@liacs.nl

**ABSTRACT**
The cyclotactor is a novel platform for finger-based tactile interaction research. The operating principle is to track vertical fingerpad position above a freely approachable surface aperture, while directly projecting a force on the same fingerpad. The projected force can be specified in Newtons, with high temporal resolution. In combination with a relatively low overall latency between tactile input and output, this is used to work towards the ideal of instant programmable haptic feedback. This enables support for output across the continuum between static force levels and vibrotactile feedback, targeting both the kinesthetic and cutaneous senses of touch. The current state of the technology is described, and an overview of the research goals of the cyclotactor project is given.

**Author Keywords**
Tactile interface, tactile interaction, haptic surface component.

**ACM Classification Keywords**
H.5.2. Information interfaces and presentation: User interfaces – Haptic I/O.

**General Terms**
Design, Human Factors.

**INTRODUCTION**
The cyclotactor is a finger-based I/O device for tactile interaction, based on three main hardware components: an electromagnet, a proximity sensor, and a "keystone" component which is attached to the fingerpad. The focus on finger-based interaction is an important difference with other magnetic haptic interfaces which have been developed under the umbrella of the Magnetic Levitation Haptic Consortium [1].

The keystone component provides both proximity input (via an infrared-reflecting surface) and force feedback (via a permanent magnet). This means that the device falls into the category of *actuated interfaces* as defined in [2]. As opposed to a tactile display, this setup allows haptic output to directly influence haptic input. In [3], where the device was introduced, it was shown how this defining characteristic can be used to set up cyclical relationships between tactile input and output, and an example of a practical application of this to musical interaction was given.

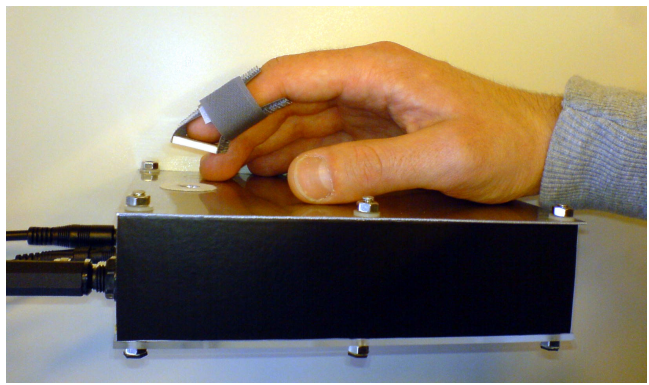

**Figure 1.** The tactile interface.

In the next Section, the current state of the technology is discussed. This is followed by an overview of the remaining research goals of the cyclotactor project, and the conclusion.

**CURRENT STATE OF THE TECHNOLOGY**
The hand, in resting position on a surface, was taken as the starting point for the current prototype. In accordance to this, the permanent magnet in the keystone was placed in a position pressed against the fingerpad (see Figure 1). This allows for natural tapping movements, and is also intended to give good transfer of vibrotactile feedback.

The host system and DSP interface hardware used are a combination of Mac OS X's CoreAudio and a Motu UltraLite, providing low-latency audio-rate voltage I/O. The output sample rate used is 192 KHz, with the associated timing precision eliminating jitter as a problem.

Custom reflected infrared (IR) detection electronics used in high-speed pulsed operation allow (after signal processing) for a vertical proximity input range of 35 mm, with a sensitivity of approximately 0.2 mm. (This represents a reconfigurable trade-off). Fingertip movement is captured at a sample rate of 4000 Hz. The system's tactile input latency is estimated at approximately 1.6 ms.

A number of electromagnets were built from scratch in order to attain a swift response for high-speed changes in coil current (and thereby magnetic output), while retaining a suitable output force range. Along with a custom opamp circuit, this currently gives an effective output frequency range of 0-1000 Hz, with a more or less linear and flat response in the vibrotactually important range up to 400 Hz (see Figure 2). The maximum added latency (to reach 99% of a magnetic target amplitude) is 1.0 ms, bringing the system's tactile output latency to an estimated 2.4 ms.

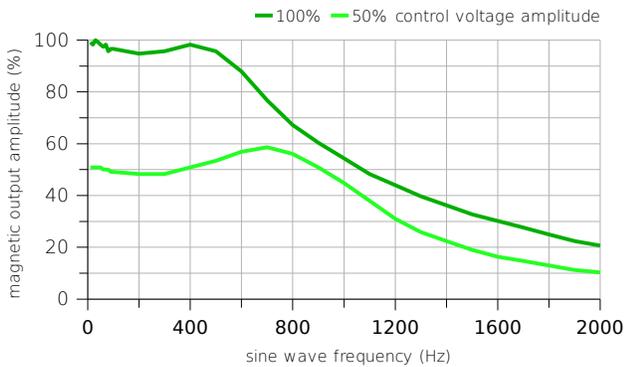

**Figure 2.** Frequency response of the magnetic output.

Keystone prototypes based on a range of permanent magnet types were created and tested. To compare different keystone/electromagnet combinations, "maximum practicable static rejection" (MPSR) was used: the maximum stable rejecting force that can be encountered across the vertical distance range, without slipping sideways of the keystone becoming unavoidable.

Since improvements to the technology were reported in [4], the system's output channel has been changed to a force signal in Newtons. The resulting force characteristic of Figure 3 underlies both kinesthetic and cutaneous feedback.

## RESEARCH GOALS

Remaining tactile research goals of the project include:

• Further perfecting the system as a platform for algorithmically expressing tactile interactions directly in terms of physical units such as Newtons, seconds, and millimeters. This in order to support the research of tactile interactions as phenomena independent of the actual technology or hardware being used.

• Developing the system as a modular surface component, creating cyclotactor arrays.

• Completing technical research already underway to remove the last obstacles in making the cyclotactor an affordable and easily replicatable platform for tactile research.

• Using the system as a tool to emulate classical tactile interactions, and, informed by them, invent new ones.

In the field of musical interaction:

• In situations of realtime musical audio synthesis where acoustical instruments are absent as predecessors, investigating the creation of completely novel controlling interactions which are however experienced as intuitive;

• Investigating musical DOFs which are created on-the-fly, "folding open" during "windows of opportunity" in interaction. For an implemented example where two virtual degrees of freedom are implemented on top of the single tactile loop, please see [2].

• Investigating the use of the above DOFs for systematically designing pathways from intuitiveness to expert performance, virtuosity, and continued exploratory musical interaction.

## CONCLUSION

The author looks forward to attending the TEI'10 GSC and meeting experienced researchers as well as students in the field of tangible, embedded and embodied interaction.

## ACKNOWLEDGMENTS

The author would like to thank Rene Overgauw at the Electronics Department of the Faculty of Science of Leiden University for his indispensable advice and practical support. Also thanks to the anonymous reviewers for their constructive criticisms. The author gratefully acknowledges the US National Science Foundation for supporting his attendance at TEI'10.

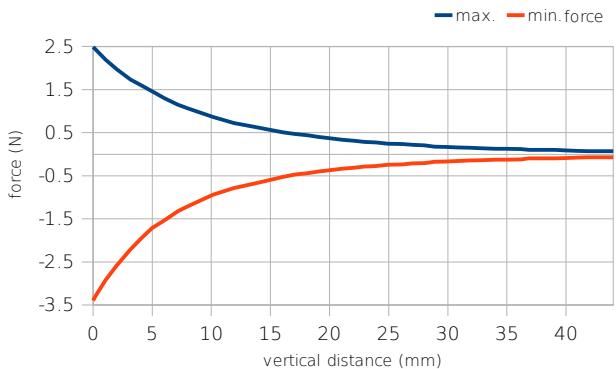

**Figure 3.** Output force range over distance above surface.